# CHARACTERIZATION OF SI HYBRID CMOS DETECTORS FOR USE IN THE SOFT X-RAY BAND

Zachary Prieskorn[a], Christopher V. Griffith[a], Stephen D. Bongiorno[a], Abraham D. Falcone[a], David N. Burrows[a]

[a]Pennsylvania State University, Dept. of Astronomy & Astrophysics, University Park, PA 16802 USA

## ABSTRACT

We report on the characterization of four Teledyne Imaging Systems HAWAII Hybrid Si CMOS detectors designed for X-ray detection. Three H1RG detectors were studied along with a specially configured H2RG. Read noise measurements were performed, with the lowest result being 7.1 $e^-$ RMS. Interpixel capacitive crosstalk (IPC) was measured for the three H1RGs and for the H2RG. The H1RGs had IPC upper limits of 4.0 - 5.5 % (up & down pixels) and 8.7 – 9.7 % (left & right pixels), indicating a clear asymmetry. Energy resolution is reported for two X-ray lines, 1.5 & 5.9 keV, at multiple temperatures between 150 – 210 K. The best resolution measured at 5.9 keV was 250 eV (4.2 %) at 150 K, with IPC contributing significantly to this measured energy distribution. The H2RG, with a unique configuration designed to decrease the capacitive coupling between ROIC pixels, had an IPC of $1.8 \pm 1.0$ % indicating a dramatic improvement in IPC with no measurable asymmetry. We also measured dark current as a function of temperature for each detector. For the detector with the lowest dark current, at 150 K, we measured a dark current of $0.020 \pm 0.001$ ($e^-$ $sec^{-1}$ $pix^{-1}$). There is also a consistent break in the fit to the dark current data for each detector. Above 180 K, all the data can be fit by the product of a power law in temperature and an exponential. Below 180 K the dark current decreases more slowly; a shallow power law or constant must be added to each fit, indicating a different form of dark current is dominant in this temperature regime. Dark current figures of merit at 293 K are estimated from the fit for each detector.

## 1. INTRODUCTION

Charge-coupled devices (CCDs) are currently the workhorse detector in the focal planes of X-ray telescopes. The technology has been in use for over 20 years and is in a high state of maturity. However, future X-ray telescopes have different requirements. Some next generation missions have up to 10x the throughput of the largest current missions (*XMM-Newton* and *Chandra*) and may also have high spatial resolution (e.g. SMART-X or Gen-X) [1], requiring detectors capable of handling an order of magnitude increase in flux density. With this kind of increase, the basic architecture of the CCD will severely limit its usefulness in future mission. CCDs will saturate at the required exposures because they cannot be read out fast enough. Budget constraints will also impose the need for the next generation of telescopes to be very long lived missions. The experience with CCDs on long missions has shown that they are vulnerable to radiation and micrometeoroid damage [2], [3]. Radiation damage results in degrading energy resolution with time and micrometeoroid damage causes blooming and/or catastrophic shorts in CCD gates, reducing the usable pixels of the CCD. While CCDs have worked well for X-ray astronomy in the past, future missions will need detectors with improved capabilities.

### 1.1. Hybrid CMOS Technology

A Hybrid CMOS detector (HCD) has an absorbing layer which can be silicon or HgCdTe depending on the application. This absorbing layer is then indium bump bonded to a read out integrated circuit (ROIC) that provides a readout electronics chain for each individual pixel in the detector (Figure 1). The advantages of this system are that the absorbing layer and ROIC can be optimized separately. The absorber is often optimized for quantum efficiency, the ROIC for improved read noise and signal processing. HCDs optimized for X-ray detection use a silicon absorber with a thin aluminum filter between the incident photons and absorber to block optical light.

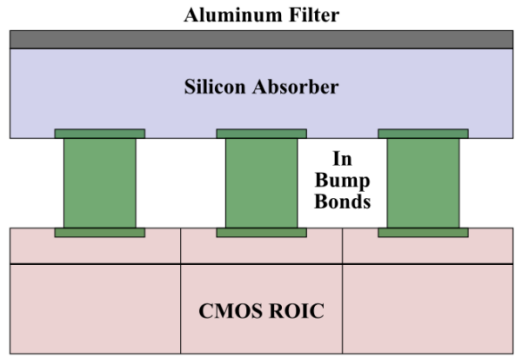

Figure 1. Schematic cross-section of an X-ray optimized Hybrid CMOS detector array. The Si absorbing layer is indium bump bonded to a read out integrated circuit (ROIC). The ROIC provides a separate read out electronics chain for each pixel in the detector. Having the two layers allows for optimization of each layer separately.

## 1.2. Hybrid CMOS Advantages for Astronomy

HCDs are already finding use in many astronomical applications in the optical/IR energy bands [4–6]. This technology will be a major improvement over CCD technology for astronomical applications and the James Webb Space Telescope will take advantage of it by flying a Teledyne Imaging Systems (TIS) infrared H2RG using a HgCdTe array absorber [7]. TIS HgCdTe HCDs have already been flight proven on the Hubble Space Telescope and WISE. TIS also produces Hybrid Visible Silicon Imagers (HyViSI), these have been launched on the Orbiting Carbon Observatory and on the Mars Reconnaissance Orbiter. The ability of HCDs to read out each pixel individually provides numerous advantages over CCDs for achieving X-ray astronomy science goals, discussed further below.

### 1.2.1. Pile-up

Pile-up occurs when multiple X-ray photons interact in the same pixel between readouts, resulting in the interpretation of multiple photons as a single event. Reducing the readout time can reduce the chance of pile-up occurring. Using the ability to read out any detector pixels, a windowed mode can be used to read out only the pixels containing interesting sources. The windowed mode allows those pixels to be read out faster. Current HCD technology allows small windows to be read out in as little time as 30 $\mu$s. Fast, full size frame reads can also be achieved using the 16 or 32 parallel readout lines for H1RGs and H2RGs respectively. This allows for the observation of very high flux sources with excellent timing information and minimal pileup effects. This ability leads to a 2-3 order of magnitude improvement in peak count rate performance over CCDs.

### 1.2.2. Radiation Damage

The direct read out of every pixel on an HCD has the added effect of making these detectors extremely radiation hard ( > 100 krads). CCDs are vulnerable to proton displacement damage because of the need to transfer charge through so much silicon (~few cm) before being read out; a problem related to the bucket-brigade read out scheme. HCDs are thus orders of magnitude less sensitive to radiation damage.

### 1.2.3. Micrometeoroids

Micrometeoroids are thought to be the cause of serious detector damage on multiple missions currently in orbit, including *XMM-Newton*, *Suzaku* and *Swift*-XRT. This micrometeoroid impact can either directly damage the CCD gate structures or indirectly affect the read out of columns through those damaged pixels. HCDs are expected to be more robust against micrometeoroid damage than CCDs. They should be protected from both failure mechanisms: the first by the fact that HCDs do not have exposed gates and the second because pixels will not bloom across the detector when damaged since each pixel has its own read out architecture.

### 1.2.4. Low Power

The on-board integration of camera drive electronics and detector signal processing reduces the power consumption and mass of HCDs compared to CCD camera designs [8]. The lower power is attributable to lower capacitance gate structures in HCDs. As an example, the Swift-XRT uses 8.4 W to produce and drive the CCD readout signals. The read out function could be achieved, even faster, with <100 mW using an H1RG HCD and SIDECAR$^{TM}$ ASIC [9]. This lower power will enable mission designs with large arrays of many small pixels operating at high rates.

## 1.3. Characterizing HCDs for use in X-ray Astronomy

Each of the following sub-sections describes the importance of our results for the continued development of HCD technology.

### 1.3.1. Energy Resolution

The energy resolution is one of the most important parameters for a new X-ray detector technology. Current CCD technology provides nearly Fano-limited energy resolution (~2.0 % at 5.9 keV). In order to capitalize on the many advantages of HCDs relative to CCDs, it is important for HCDs to approach this Fano-limit. Energy resolution in HCDs is currently limited by interpixel capacitance crosstalk (IPC) and read noise.

### 1.3.2. Interpixel Capacitance Crosstalk

Interpixel capacitance crosstalk (IPC) is an artificial presence of signal in the pixels surrounding a pixel which had a true photon-induced signal. The effect is observed most strongly in the pixels immediately adjacent to the central pixel and to a lesser extent in the diagonal pixels. Values for IPC reported in this paper are of the pixels adjacent to the primary pixel, either directly up, down, left or right. IPC is believed to be caused by capacitive coupling which occurs between neighboring absorber and/or ROIC pixels; a schematic is shown in the left panel of Figure 2. The major impact of IPC to X-ray applications is a reduction in the energy resolution and a smearing of the image (Figure 2, right panel). The energy resolution is degraded both by the fact that spurious signal is introduced to the system and by increased noise due to reading out multiple pixels for a single event. Minimizing IPC is one of the most important steps for current HCD development focused on X-ray detection.

### 1.3.3. Read Noise

The read noise sets the noise floor for a detector read out. It is caused by the various steps involved in processing the signal from the point where it enters the ROIC until it is read out by a computer as an A/D channel number. A higher noise floor degrades the energy resolution. At slow read out speeds the read noise in HCDs is currently worse than that found in the best CCDs. However, at Megapixel/s read out speeds, HCDs have lower read noise than CCDs [4].

### 1.3.4. Dark Current

Dark current is charge observed in the detector when it is not exposed to any radiation source. The most common causes are from defects in the bulk silicon which trap charge and then release it over time, and from defects at the Si/SiO$_2$ interface. This thermal charge is overwhelming at room temperatures for X-ray detectors; thus these detectors must be cooled. The expense of cooling detectors for flight missions can be prohibitive so it is important to optimize the operating temperature for minimal dark current.

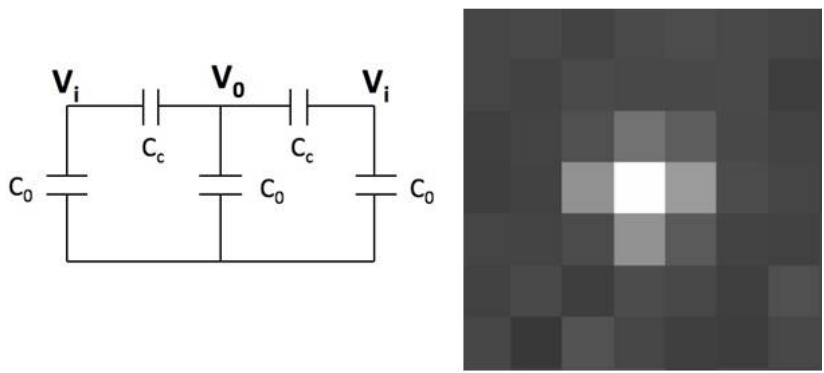

Figure 2. (Left) A schematic of the interpixel capacitance crosstalk (IPC). $V_o$ is the central pixel voltage, $V_i$ is the voltage of neighboring pixels, $C_c$ is the coupling capacitance between neighboring pixels and $C_o$ is the expected capacitance between the pixel and readout node. There would also be crosstalk between pixels into and out of the page. (Right) An image from one of the H1RG detectors showing significant IPC. The event is from a 5.9 keV X-ray. Signal should only show up in the central pixel but because of IPC has spread to the surrounding pixels.

## 2. EXPERIMENT SETUP

### 2.1. PSU HCD Test Stand

We report results on four different HxRG[1] detectors tested using an experimental setup at PSU. In each case, the detector package is mounted inside a light-tight stainless steel vacuum chamber, shown in Figure 3 and Figure 4. To prevent condensation on the cold detector surface and to minimize X-ray attenuation, this chamber is evacuated to a pressure of less than 10$^{-5}$ Torr before the detector is cooled. The detector is mounted to a cold finger connected to a liquid nitrogen (LN2) reservoir outside the chamber via a feedthrough in the chamber wall. The LN2 flow is controlled by a solenoid valve operated by a National Instruments LabVIEW 10 program. The temperature was successfully controlled between 150-210 K with a precision of ±0.2 K.

The detector package is connected to a room temperature SIDECAR$^{TM}$ controller via a 92 line flex cable. The SIDECAR$^{TM}$ provides clock and bias signals to the detector while performing chip programming, signal amplification, analog to digital conversion and data buffering. The SIDECAR$^{TM}$ is mounted to the room temperature chamber wall inside the vacuum chamber. Signal processing and further amplification are performed with the TIS JADE-2 card. This board is mounted to the SIDECAR$^{TM}$ electronics board and passes the fully amplified signal out of the vacuum chamber via a USB 2.0 feedthrough in the chamber base.

The detector readout is controlled with the TIS JAC software package. This program allows the user to control many aspects of the detector readout. In our data taking we primarily control the number of images being collected and the reset scheme. A single data run involves resetting the HxRG at the beginning followed by a "ramp" of image taking. During a ramp, charge is read out every 5.28 s and stored as an image, but the detector is not reset, leaving the charge in the pixels. At the end of each ramp, the detector is reset, clearing the charge before a new ramp begins. To separate the signal from fixed pattern noise and dark current, a software correlated double sample (CDS) is performed, described in Sec. 3.

---

[1] "HxRG" will be used to refer to a broad class of any sized HAWAII silicon CMOS detector.

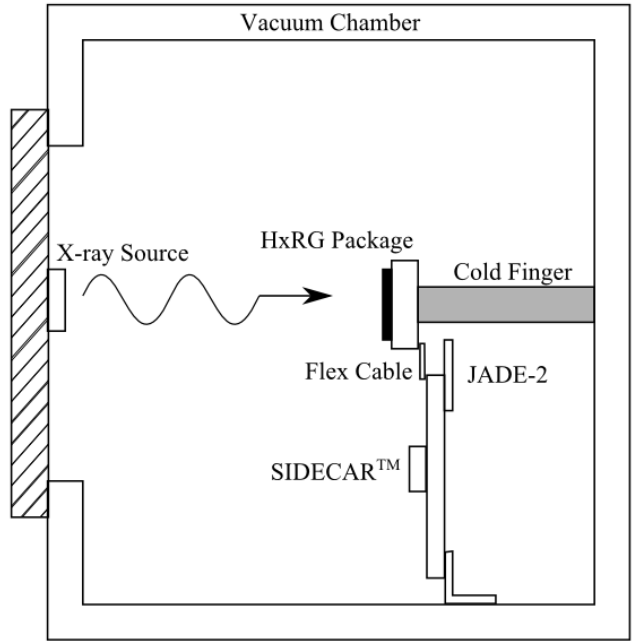

Figure 3. Schematic of the inside of the PSU HCD test stand. The HxRG package is mounted to a cold finger which cools the detector using LN2 provided via a feedthrough in the chamber wall. The detector is connected to a SIDECAR$^{TM}$ using a 92 line flex cable. The SIDECAR$^{TM}$ is mounted to the room temperature base of the chamber. An X-ray source is mounted opposite the HCD. Sec. 2.1 provides a more detailed description of the experiment setup.

Two X-ray sources were used to characterize the energy resolution of the HxRG's in this experiment. Two different source plates can be mounted to the vacuum chamber. The first plate has an $^{55}$Fe source mounted to it. This source produces Mn K$\alpha$ (5.9 keV) and K$\beta$ (6.4 keV) lines. A shutter is manually controlled from outside the chamber and can fully block the source from the detector. The second plate configuration has two $^{210}$Po sources mounted above a target material. Alpha particles from the $^{210}$Po fluoresce the target material, producing X-rays of a characteristic energy for that material. We used only Al (1.5 keV K$\alpha$) as a fluorescent material in this experiment.

2.2. Detectors

The Hybrid CMOS detectors studied in this experiment were all produced by TIS. Hybrid CMOS technology is described in Sec. 1.1. The specific detector characteristics for this project are presented in Table 1. The first set of detectors tested were Hawaii-1RG (H1RG) type detectors, where the "1" signifies a 1024x1024 pixel configuration. Each of the H1RG detectors have a pixel pitch of 18 $\mu$m. A specially modified engineering grade H2RG detector was also tested. This detector has a 2048x2048 read out integrated circuit (ROIC) with 18 $\mu$m pitch pixels indium bump bonded to a 1024x1024 absorber layer with 36 $\mu$m pitch pixels. Only one ROIC pixel is bonded to each absorber pixel, creating an effective 36 $\mu$m pitch for the ROIC pixels.

The HxRG detectors were originally optimized for Optical/IR astronomy. These detectors have reached a high technology readiness level through the Mars Reconnaissance Orbiter and Orbiting Carbon Observatory programs but these devices have not been optimized for X-rays [7]. The first change in these detectors for X-ray optimization was to remove the anti-reflection coating, and to instead provide an aluminum optical blocking filter. The H1RGs all have aluminum optical filters of varying thicknesses, reported in Table 1. An optical filter is typically required for detectors on X-ray telescopes because light from stars and other background sources will otherwise contaminate the X-ray images but the optical blocking filters are typically deposited on a substrate that is mounted separately from the X-ray detector.

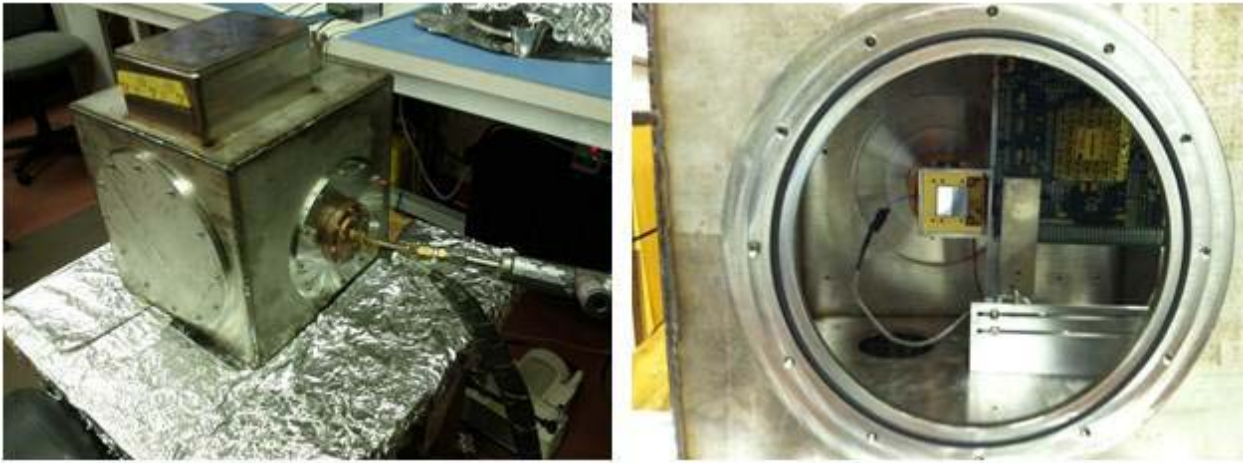

Figure 4. PSU test stand images. The left panel is the vacuum chamber showing liquid nitrogen feedthroughs connected to a cold finger inside the detector. The right panel shows the chamber inside with a Hybrid CMOS detector attached to the cold finger and the SIDECAR$^{TM}$ to the right of the detector.

The filters on the detectors in this experiment were all deposited directly to the surface of the detector but, in some cases, only on one half the detector surface. This allowed us to compare the detector response with and without the optical filter. Transmission curves at wavelengths from 100-1000 Å for the different aluminum filters are reported in [9].

During testing of the detector H1RG-125 in 2010, the detector was accidentally exposed to bright optical light with power applied to $V_{sub}$, causing deep saturation that resulted in a permanent threshold shift in the unfiltered half of the detector. The effect decreases dynamic range and produces excess noise across the entire chip [10]; therefore we only include events from the aluminum coated side for measurements of IPC and dark current. The pre-damage energy resolution is also reported.

Table 1. Hybrid CMOS detector parameters

|  | Absorber Pitch ($\mu$m) | ROIC Pitch ($\mu$m) | Absorber Dimensions (pixels) | ROIC Dimensions (pixels) | Optical Filter Thickness (Å) |
|---|---|---|---|---|---|
| H1RG-125 | 18 | 18 | 1024x1024 | 1024x1024 | 500 |
| H1RG-161 | 18 | 18 | 1024x1024 | 1024x1024 | 1000 |
| H1RG-167 | 18 | 18 | 1024x1024 | 1024x1024 | 180 |
| H2RG-122 | 36 | 36(18)[a] | 1024x1024 | 2048x2048 | 0 |

[a]The ROIC pitch for the H2RG was 18 $\mu$m but only one ROIC in every four is bonded to each 36 $\mu$m absorber pixel. This leads to an effective ROIC pitch of 36 $\mu$m.

## 3. ANALYSIS AND RESULTS

Using the setup described in Sec. 2.1, data were collected for each detector. While the data taking process differs depending on the measurements being made – IPC (Sec. 3.2), energy resolution (Sec.3.3), or dark current (Sec. 3.3.3) – there are some event grading steps in common for all three (Sec. 3.1). The X-ray images are acquired using the JAC and its software, as described in Sec. 2.1. The ramped images collected are subtracted in software using a script. All image analysis was performed with code written in

IDL. The image subtraction process is referred to as CDS. Each image is subtracted from the image immediately following it in time. This removes the fixed-pattern noise, which should be the same from image to image. After the CDS we are still left with horizontal artifacts that we call "row noise". To remove this effect we subtract from each row the output of a 15 pixel moving median filter [10]. This leaves a single image with clear X-ray events and a noise floor at the read noise level of the detector plus a dark current signal. The details of the analysis specific to each experiment are described below.

### 3.1. Event Grading

After the software CDS process we perform X-ray event detection. Each image is scanned with a single pixel threshold referred to as the primary threshold. Any pixel found with an analog-to-digital converter signal value – digital number (DN) – above this threshold is saved in an event list with the 3x3 pixel neighborhood centered on the above threshold pixel. To be included as an event, the primary pixel must also be the local maximum within the event neighborhood .

For the detector with low IPC (H2RG-122) we use the following technique for event grading. The event list is further filtered by grading the events using the *Swift*/XRT grade definitions [11]. A secondary threshold is applied to each event neighborhood. Any pixel in the neighborhood above the secondary threshold is included in the graded X-ray event. This threshold is typically set a few sigma above the noise floor; see Sec. 3.3 for a description of how the noise floor is determined. Each event is graded by the number and position of pixels in the neighborhood that exceed this secondary threshold. Grade 0-4 events are the only ones included in the energy resolution measurements reported in Sec.3.3. Grade 0 events have zero pixels above the secondary threshold in addition to the central pixel. Grade 1-4 events are commonly referred to as "split" events and have a single pixel above the secondary threshold besides the central pixel. This pixel is located in one of the positions directly up, down, left or right from the central pixel. The event energy is then the sum of the pixels in the event neighborhood that exceed the secondary threshold.

For detectors with high IPC (H1RGs), a different method for determining event grades must be used due to the large amount of IPC signal in pixels surrounding the pixels with signal directly related to the X-ray event. The fraction of the central pixel signal contained in the second brightest pixel in the 9 pixel event neighborhood is calculated. This quantity is referred to as the event split. An upper and lower threshold of the event split are used to select only events likely to be Grade 0-4, if the IPC were removed. Given the high IPC, these events all appear as cross-like. A histogram of the total events in the event split is used to set the threshold for including events in the energy resolution and IPC calculations.

### 3.2. Interpixel Capacitance Crosstalk

#### 3.2.1. Analysis

IPC is reported for three separate techniques in this paper. We measured the IPC of the CMOS detectors by finding the most symmetric X-ray events in the $^{55}$Fe source data at 150 K. To effectively measure IPC, we did not want to include "split" events. "Split" events occur when an X-ray charge is split between two or more pixels due to the position on the pixel where the X-ray is detected. If the X-ray charge cloud diffuses into two or more pixels as it reaches the base of the pixel, the charge will be split between them. If these events are included in the IPC measurements they skew the result because spreading of the charge in "split" events is not due to IPC. IPC values for H1RGs are quoted as upper limits because the potential charge diffusion throughout a pixel with 100 $\mu$m thick absorption layer could have a FWHM of approximately 15 $\mu$m with a bias voltage of 15 V [7]. This is probably an overestimate of the charge diffusion width since the full 100 $\mu$m depth is not seen by the charge cloud from an X-ray that interacts below the surface. 15 $\mu$m of diffusion is less than the 18 $\mu$m pitch of the H1RG detectors but it is close enough that some charge spreading may be contributing to our IPC measurements so we conservatively quote them as upper limits. By selecting only symmetric cross-like events, we make the best possible estimate of the IPC for H1RGs. The larger pixel size of the modified H2RG (36 $\mu$m) allows the IPC to be accurately calculated without charge diffusion effects.

Events detected via a "primary threshold" selection must satisfy the following criteria to be counted as X-rays. We first determined whether an event was a split event by using a "secondary threshold". This criterion looks at the surrounding eight pixels of the X-ray event and if one pixel is measured to be above the threshold, measured in DN, it will graded as a split event and therefore eliminated from the sample of events used for IPC calculation. We next used a region cut to avoid using X-ray events near the edge of the detector where measured gain variation is present. X-ray events were only included if the second brightest pixel in the X-ray event was ≤ 40% the value of the center pixel, since they would otherwise be clear split events. The total signal of all nine pixels in the event was measured to verify the event was from an $^{55}$Fe X-ray. We found the means of the manganese K$\alpha$ and K$\beta$ peaks and set thresholds 4$\sigma$ below the K$\alpha$ peak and 4$\sigma$ above the K$\beta$ peak to ensure all of the symmetric $^{55}$Fe X-ray events were being used.

Since the above cuts will still include some split events, we next used a standard deviation method to find the most symmetric events since split events will not be symmetric (an X-ray cannot simultaneously land on more than one side of a pixel). We first measured the standard deviation of the "nearest neighbor" pixels of the event, those pixels directly up, down, left, and right of the central pixel. For an event to be included in what we termed the "paired pixel" method the standard deviation of the (up/down) pixels needed to be less than 1$\sigma$ of the value of the noise floor and the standard deviation of the (left/right) pixels also needed to be less than 1$\sigma$ of the value of the noise floor. This technique is sensitive to any asymmetries between the (up/down) and (left/right) pairs of pixels. The results of this technique are reported in Sec. 3.2.2.

Our previous studies have used two different methods for measuring IPC. One previous method characterized IPC by including only events that contain a second brightest pixel in the event neighborhood [10]. This method has the advantage of being sensitive to asymmetries between (up/down) and (left/right) paired pixels. IPC values of ~5 % (up/down) and ~7% (left/right) were reported for the H1RG detectors in this study. The down side to this method is the inclusion of too many split events.

Another method used is similar to the "paired pixel" method described in this paper, but compares all 4 "closest neighbor" pixels to each other [12]. This method is efficient at removing split events but preferentially selects only the most symmetric events, missing potential asymmetries in the IPC distribution. This study found IPC values of ~6.5-8 % (up/down) and ~6.5-8 % (left/right).

3.2.2. Results
We report IPC results here using the "paired pixel" technique described in Sec. 3.2. This method shows any asymmetries between the (up/down) and (left/right) pixels. The three H1RG detectors tested in this experiment have the standard Hybrid CMOS pixel configuration, with one absorbing 18 $\mu$m pixel bonded to an 18 $\mu$m ROIC pixel. These detectors have been found to have large IPC, as discussed in Sec.1.3.2. The IPC results from these three detectors are reported in Table 2. The table for each detector shows the 9 pixel neighborhood surrounding an X-ray event. The value in each cell is the percentage of the total event signal. The total event signal is the sum of the signal in all nine pixels. As described in Sec. 3.2, only X-ray events with symmetric signal were used to measure IPC. The H1RGs have IPC upper limits of 4.0 – 5.5 % (up/down) and 8.7 – 9.7 % (left/right) of the total signal measured in the event neighborhood . For all three H1RGs there is a clear asymmetry, indicating a different IPC between pixels of the same row and pixels in the same column.

An engineering model H2RG was specially constructed by TIS to reduce the IPC, as described in Sec.

2.2. If the coupling capacitance is occurring between the pixels as suggested by [13], then the extra distance between pixels should reduce the capacitance between active pixels and thus reduce the amount of signal detected in adjacent pixels. Table 2, lower right quadrant, reports the IPC measurements for H2RG-122. The average IPC of the four adjacent pixels to the central pixel is 1.8 ± 1.0 % of the total neighborhood signal which is approximately 2 % less than the (up/down) pixels and 7 % less than the (left/right) pixels of the best H1RG detector, H1RG-161. The modified H2RG is a significant improvement over the standard Hybrid CMOS pixel configuration regarding IPC and does not show an asymmetry between (up/down) and (left/right) pixels.

Table 2. IPC measurements using the Standard Deviation (Paired Pixel) technique for the four detectors in our study. The H1RG values reported are all upper limits, see Sec. 3.2. Each H1RG shows a clear asymmetry between the (up/down) and (left/right) pairs of pixels. H2RG-122 IPC values are not upper limits and do not show measurable asymmetry.

H1RG-125

| 0.017 ± 0.007 | 0.054 ± 0.010 | 0.018 ± 0.008 |
|---|---|---|
| 0.097 ± 0.013 | 0.626 ± 0.068 | 0.097 ± 0.013 |
| 0.018 ± 0.007 | 0.054 ± 0.010 | 0.018 ± 0.007 |

H1RG-161

| 0.011 ± 0.008 | 0.044 ± 0.014 | 0.012 ± 0.008 |
|---|---|---|
| 0.087 ± 0.015 | 0.690 ± 0.098 | 0.087 ± 0.015 |
| 0.011 ± 0.009 | 0.045 ± 0.013 | 0.011 ± 0.007 |

H1RG-167

| 0.012 ± 0.006 | 0.040 ± 0.007 | 0.013 ± 0.005 |
|---|---|---|
| 0.093 ± 0.010 | 0.685 ± 0.062 | 0.093 ± 0.010 |
| 0.012 ± 0.004 | 0.039 ± 0.006 | 0.013 ± 0.005 |

H2RG-122

| 0.007 ± 0.012 | 0.017 ± 0.010 | 0.007 ± 0.011 |
|---|---|---|
| 0.018 ± 0.010 | 0.901 ± 0.11 | 0.018 ± 0.011 |
| 0.007 ± 0.012 | 0.017 ± 0.010 | 0.007 ± 0.012 |

3.3. Energy Resolution

3.3.1. Analysis

After the events were graded, an energy spectrum was made using all of the events that pass the primary and secondary threshold cuts (as defined in Sec. 3.1). The energy resolution was calculated by fitting peaks in the energy spectra with Gaussian distributions. For the $^{55}$Fe spectra, Gaussians were fit to the noise peak and a Gaussian for the Mn K$\alpha$ and K$\beta$ lines. The Al spectra is fit with Gaussians for the noise peak, the Al K$\alpha$ line, and a Si K$\alpha$ line. The Si is from the Al alloy used as a target. The full-width-half-maximum (FWHM) or $\Delta E$, where $E$ is the energy centroid of the event from the Gaussian fit, value of the distribution is then given as $\Delta E = 2\sqrt{2 \ln 2}\,\sigma$, where $\sigma$ is the standard deviation found from the Gaussian fit to the line of interest.

3.3.2. Results

An $^{55}$Fe source was used to measure the energy resolution at 5.9 keV for each detector. An Al line produced by a fluorescent source was used to measure energy resolution at 1.5 keV. Energy resolution was measured at multiple temperatures between 150 – 210 K. H1RG-125 had the best energy resolution, 248 eV at 5.9 keV (4.2 %) in measurements made in 2008 [10]. As discussed in Sec. 2.2, H1RG-125 was damaged after the 2008 experiment and was thus not available for further testing in this experiment. The energy resolution measurements for each detector are reported in Table 3 and vary from 285 – 539 eV at

5.9 keV (150 K) and 156 – 266 eV at 1.5 keV (150 K). Spectra for the Al fluorescent source and the $^{55}$Fe source are shown in Figure 5. The fit for each spectrum is shown as a solid line.

The modified H2RG detector has much improved IPC compared to the H1RGs. This allows us to perform event grading in a manner similar to CCDs, as described in Sec. 3.1. We report an energy resolution of 443 eV at 5.9 keV (7.5 %) for this detector using grade 0-4 events, shown in

Figure 6 (Left). Using only grade 0 events (single pixel) we expect the energy resolution to improve as the result of fewer pixels being included in the energy sum for each event. At 150 K, we measured an energy resolution of 369 eV at 5.9 keV (6.3 %) using only grade 0 events for H2RG-122 and this improved spectrum is shown in

Figure 6 (right). The ability to select only grade 0 events, resulting in a significant improvement in energy resolution, is a demonstration of this detector's improved IPC characteristics. It should be noted that the improved energy resolution is limited by the fact that H2RG-122 is an engineering grade device.

Table 3. Energy resolution as a function of temperature and energy for TIS HCDs.

| | H1RG-125[a] | | | | |
|---|---|---|---|---|---|
| Line Energy (keV) | 150K (eV) | 160K (eV) | 170K (eV) | 180K (eV) | 190K (eV) |
| **5.9 (Mn K$\alpha$)** | 248 | - | - | - | - |
| | H1RG-161 | | | | |
| Line Energy (keV) | 150K (eV) | 160K (eV) | 170K (eV) | 180K (eV) | 190K (eV) |
| **5.9 (Mn K$\alpha$)** | 539±44 | 506±9 | 493±7 | 568±8 | 606±10 |
| **1.5 (Al K$\alpha$)** | 266±9 | 264±6 | 360±20 | 352±14 | 429±12 |
| | H1RG-167 | | | | |
| Line Energy (keV) | 150K (eV) | 160K (eV) | 170K (eV) | 180K (eV) | 190K (eV) |
| **5.9 (Mn K$\alpha$)** | 285±2 | 310±4 | 303±3 | 329±5 | 331±3 |
| **1.5 (Al K$\alpha$)** | 210±10 | 229±12 | 236±11 | 226±11 | 299±19 |
| | H2RG-122 (Grade 0 Events)[b] | | | | |
| Line Energy (keV) | 150K (eV) | 160K (eV) | 170K (eV) | 180K (eV) | 190K (eV) |
| **5.9 (Mn K$\alpha$)** | 369±4 | - | 384±2 | - | 468±5 |
| **1.5 (Al K$\alpha$)** | 156±8 | - | 182±17 | - | 344±12 |

[a]Data for H1RG-125 were not obtained for any temperatures other than 150 K.
[b]Data for H2RG-122 were not obtained at 160 or 180 K.

The event grading used for the energy resolution measurement for each H1RG detector must be adjusted as a function of energy because larger real X-ray related signal leads directly to higher IPC-related signal in the surrounding pixels. This adjustment is accomplished by changing the secondary threshold. While this decreases the robustness of the energy resolution measurements, we observe a trend in the secondary threshold as a function of energy. The secondary threshold that minimizes the energy resolution decreases with energy for all four detectors, from approximately 60 DN at 5.9 keV to approximately 20 DN at 1.0 keV. The decreasing threshold is set at 5.9 keV and then scaled based on the IPC contribution to the measurement for lower energies, thus the scaling is not arbitrary.

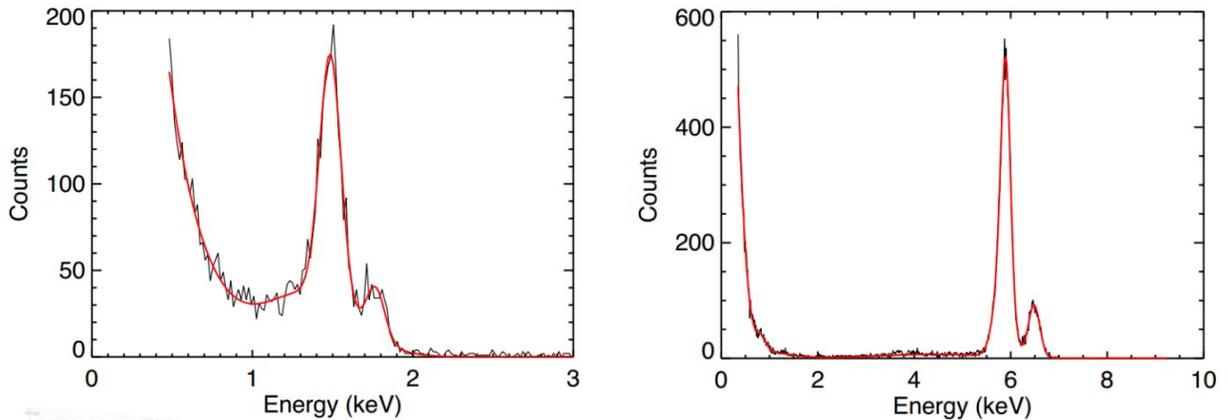

Figure 5. (Left) An Al spectrum taken with H1RG-161 at 150 K showing the fit as a solid red line. The fit includes Gaussian components for the noise peak, the Al K$\alpha$ line at 1.5 keV, and a contaminating Si K$\alpha$ line at 1.7 keV. The energy resolution of the Al line is 266 eV (17.7 %). (Right) An $^{55}$Fe spectrum taken with H1RG-167 at 150 K. The fit includes Gaussian components for the noise peak, the 5.9 keV (Mn K$\alpha$) line, and the 6.4 keV (Mn K$\beta$) line. The energy resolution of the 5.9 keV line is 285 eV (4.8 %).

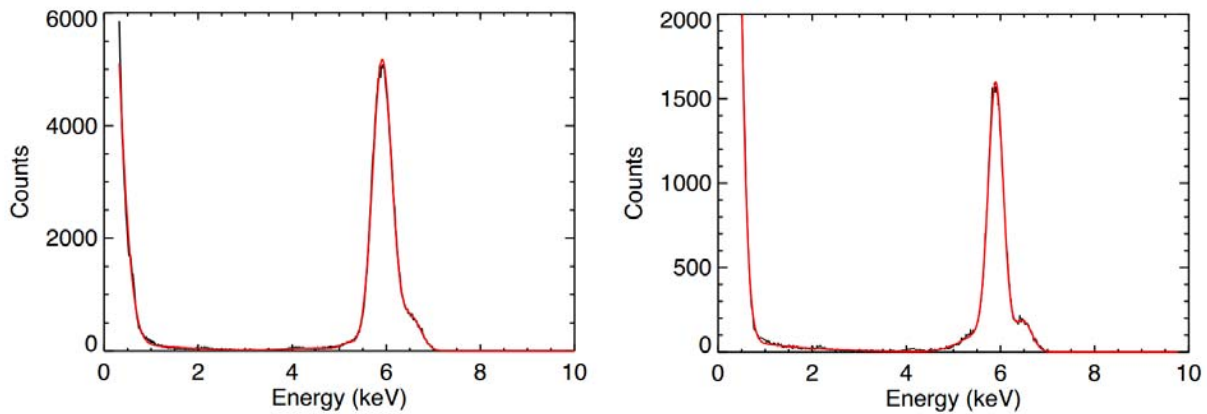

Figure 6. (Left) $^{55}$Fe spectrum taken with H2RG-122 at 150 K including all grade 0-4 events, as described in Sec. 3.1. The energy resolution of the 5.9 keV line (Mn K$\alpha$) is 443 eV (7.5 %). (Right) $^{55}$Fe spectrum taken with H2RG-122 but only including grade 0 events. The energy resolution is improved to 369 eV (6.3 %). The capability of selecting only grade 0 events is only possible due to the low IPC characteristics of this detector.

Plots of the pixels/event for each detector's Fe$^{55}$ energy spectra are shown in Figure 7. The H1RGs all have between 5 – 6 pixels for most events included in the energy spectra. H2RG-122 includes 2 pixels in most events in its energy spectrum, a clear indication of the reduced IPC and/or charge spreading in this detector compared with the 18 $\mu$m H1RGs in our experiment. This smaller number of pixels with measurable signal from an X-ray event provides the ability to characterize the X-ray event energy by reading only one or two pixels rather than ~5 pixels. Similar results are observed with the aluminum spectra.

The temperature dependence of energy resolution was measured and is shown for each detector in Figure 8 (left). The energy resolution remains mostly unchanged within error bars below approximately 170 K for each detector, indicating the possibility of running the HCDs at temperatures greater than 150 K without significant impact on detector performance, even for slow readout speeds.

Figure 8 (right) shows plots of the energy resolution as a function of energy for each detector (H1RG-161, H1RG-167, and H2RG-122) at 150 K. We fit the two points for each plot with a power-law distribution of the form $\Delta E \propto E^{1/2}$, shown as a solid red line. For a typical Si charge coupled device, this dependence on energy should be roughly applicable above the energy where read noise dominates the signal. However, there are additional effects on $\Delta E$ due to IPC, which confound the direct application of this relationship to these detectors. In spite of this, the measured energy resolution does approximate the expected form for all the detectors with some fluctuations above and below.

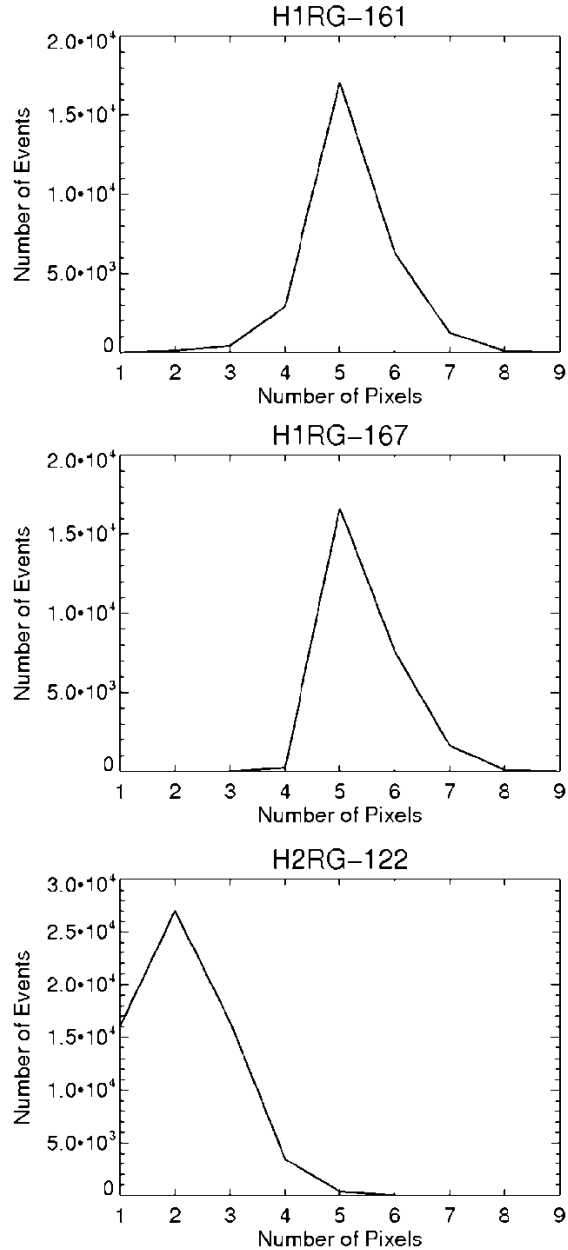

Figure 7. Plots of the number of pixels included in the $^{55}$Fe events which make up the energy spectra at 150 K using a secondary threshold of 60 DN for each detector. In the H1RG detectors, the energy spectra events primarily contain between 5 – 6 pixels, clearly indicating the effects of IPC spreading the charge across the event island . The events included in the energy spectrum of detector H2RG-122 peak around 2 pixels/event, due to improved IPC characteristics of this detector. The plots for Al have the same characteristics.

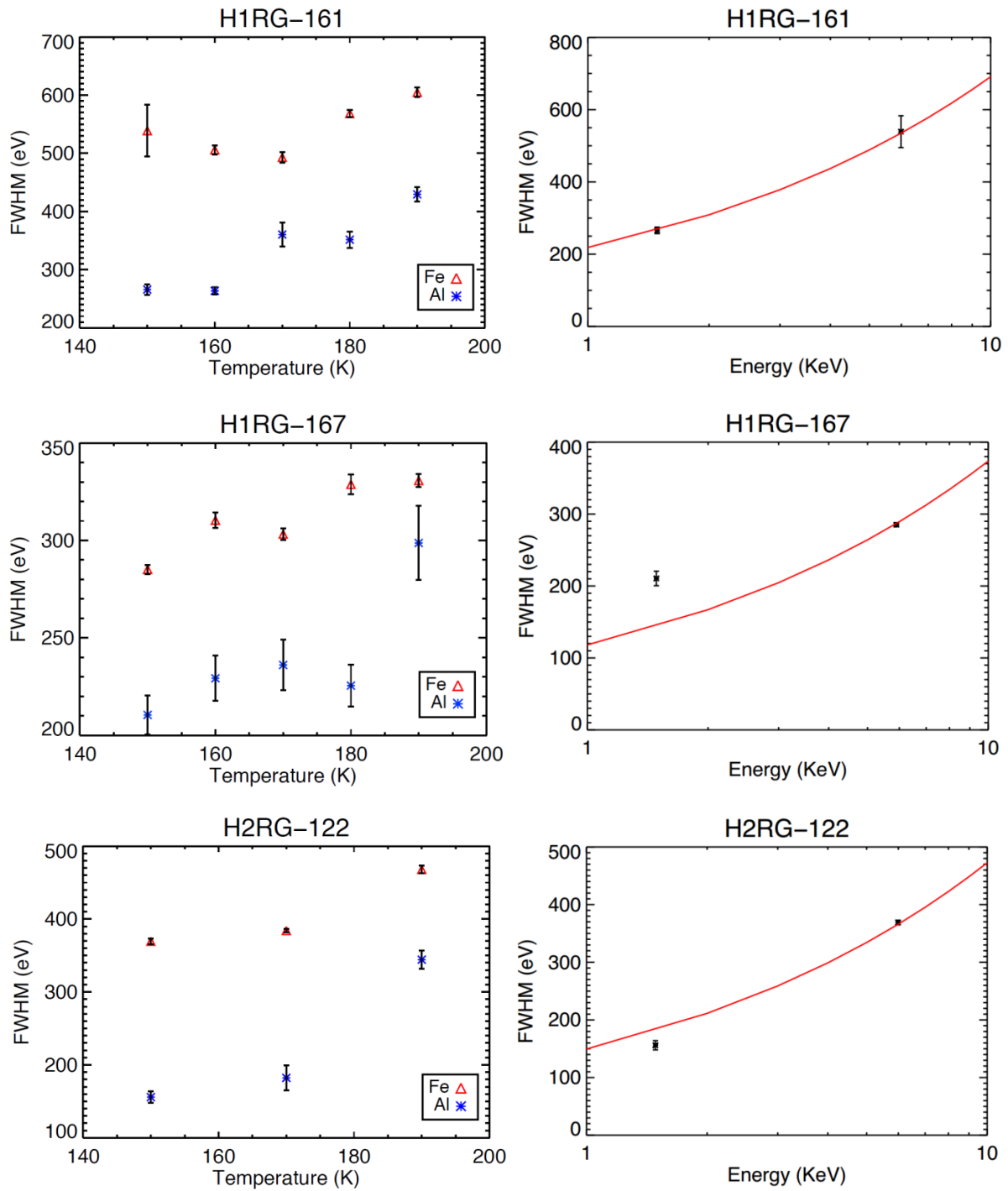

Figure 8. (Left) The left column of plots show the energy resolution as a function of temperature for each detector. The red triangles are for 5.9 keV lines from the $^{55}$Fe source. The blue asterisks are for 1.5 keV from the Al source. The temperature dependence of the energy resolution decreases below approximately 170 K, indicating the possibility of running these detectors above 150 K without significant impact on performance. (Right) These plots show the energy resolution as a function of energy for each detector. The solid red line is a fit to the two points of the form $\Delta E \propto E^{1/2}$ and they show that the measurements are roughly consistent with the expected form.

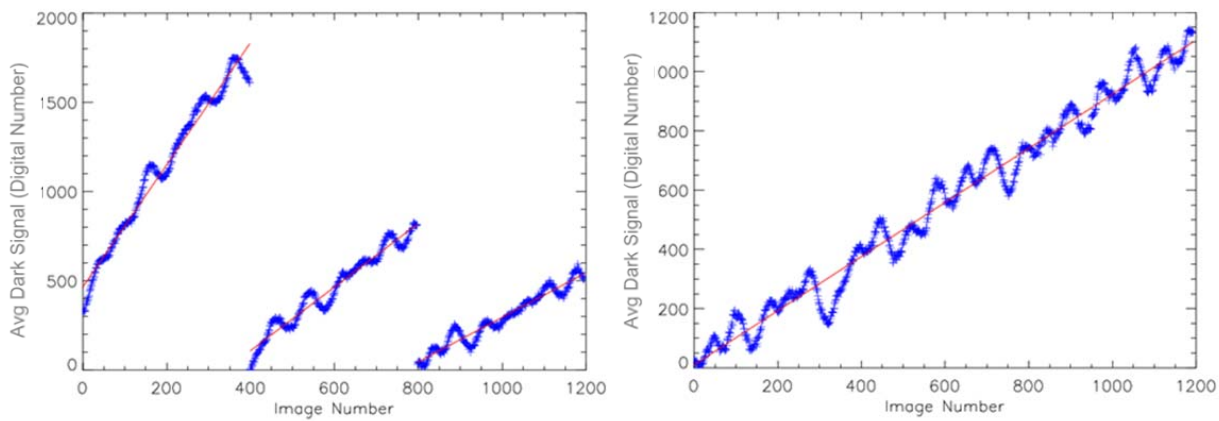

Figure 9. Measurements of dark current made using detector H1RG-161. The left panel shows the average (DN) of 3x400 image ramps taken after the detector had been cooled down to 160 K. The small scale variations are due to small temperature fluctuations but there is also a larger change in dark current happening over the ~120 min required to complete all three ramps, clear from the changing slope in the left panel (the slope in each ramp varies from 1.0-3.4). The right panel is another set of 3x400 image ramps taken immediately after the data set in the left panel and has a slope of 0.91. The ramps have been tied end-to-end to improve the linear fit. The data taken in the right panel have reached an equilibrium point and become consistent. The data in the right panel are also consistent with the last ramp in the left panel, indicating that the equilibrium point is reached approximately 80 min after the detector begins taking data.

### 3.3.3. Read Noise

The read noise was measured using the average of the pixel-by-pixel standard deviation of 200 dark exposures at 150 K for each detector. Assuming that the dark current at 150 K is negligible compared to the read noise level, this gave us the read noise in units of DN. The electron/DN conversion was found by using an estimate of 1616 electrons per event (assuming $w$ = 3.65 eV/e$^-$ and a 5.90 keV X-ray) for the Mn K$\alpha$ line and by finding the mean of the Mn K$\alpha$ peak in each 150 K spectrum. This conversion factor was then used to calculate the read noise in units of electrons. The read noise is reported in Table 4, values varied from approximately 7 – 16 e$^-$.

## 3.4. Dark Current

### 3.4.1. Analysis

Dark current images were taken with the ramp method described at the beginning of Sec. 3, however no X-ray source was on while the images were taken. Dark current data were taken at multiple temperatures for each detector, from 150-210 K. Ramps containing 400 images each were taken so that the dark current could be clearly distinguished from the noise floor. The ramp images were processed as described in Sec. 3.1, without the software CDS, allowing the dark current signal to add up image to image. Since there is no X-ray signal, the read noise and dark current are the dominant sources of signal in each image. We assume the read noise is constant, with some small Poisson fluctuation, from image to image in the ramp. Comparing the observed dark signal from successive ramp images should reveal the amount of dark current present in the detector at a given temperature.

To determine the signal in each image we take the average of all the pixels in the image except for those within 10% of the edge of the detector. The outer region of the detector is avoided because of potential gain variation around the edge. The average is then plotted for each image and fit by a linear function. The slope of this function is the detector dark current at a given temperature. Each ramp was fit independently and averaged together to produce the final dark current value.

We chose to fit the dark current data as a function of temperature with the theoretical dark current function given below and described in [14]:

$$DC(e-/sec) = 2.5 \times 10^{15} P_s D_{FM} T^{1.5} exp\left(\frac{-E_g}{2kT}\right) \qquad (1)$$

where $P_s$ is the pixel size in cm$^2$, $T$ is the temperature (K), $D_{FM}$ is the dark current figure of merit (e$^-$ sec$^{-1}$ cm$^{-2}$) at $T = 293$ K, $k$ is Boltzmann's constant, and $E_g$ (eV) is the silicon band gap energy as a function of $T$ and is defined as the following:

$$E_g = 1.1557 - \frac{7.021 \times 10^{-4} T^2}{1108 + T} \qquad (2)$$

We also fit the data with variations of this function that included the addition of a constant, a shallower power law and another dark current function with different $D_{FM}$ values.

3.4.2. Results

Dark current measurements were made for four TIS HxRG detectors at temperatures between 150 – 210 K. Most of the data used were from 3x400 image ramp data sets. Each 400 image ramp takes approximately 40 min to complete. The ramps used in our analysis were only included after two charge clearing ramps had been run. We found that charge clearing ramps are necessary before collecting dark current images. The dark current was observed to decrease over time until becoming consistent after approximately 80 minutes of continuous data collection. Figure 9 shows the observed effect and the more consistent data after a charge clearing ramp.

Table 4. Dark current, read noise measurements and average IPC for each detector

|  | 150 K, Data (e$^-$/sec/pixel) | 150 K, Teledyne (e$^-$/sec/pixel) | 293 K, Fit[a] (e$^-$/sec/pixel) | Read Noise (e$^-$) (RMS) | Avg IPC Up/Down (%) | Avg IPC Left/Right (%) |
|---|---|---|---|---|---|---|
| H1RG-125 (filtered) | 0.280 ± 0.080 | 0.284 | 3.78E6 | 10.37 | 0.054 ± 0.010 | 0.097 ± 0.013 |
| H1RG-125 (unfiltered) | 0.230 ± 0.023 | 0.284 | N/A | N/A | N/A | N/A |
| H1RG-161 | 0.020 ± 0.005 | 0.007 | 8.15E5 | 10.64 | 0.045 ± 0.014 | 0.087 ± 0.015 |
| H1RG-167 | 0.056 ± 0.026 | 0.069 | 4.45E6 | 7.05 | 0.040 ± 0.007 | 0.093 ± 0.010 |
| H2RG-122 | 0.020 ± 0.001 | N/A | 9.38E6 | 16.31 | 0.017 ± 0.010 | 0.018 ± 0.010 |

[a]These values are from a fit to the dark current data using Equation 1

We plot the dark current at 150 K for each detector and compare to the values reported by TIS for each detector (Figure 10). The PSU measurements are roughly consistent with those from TIS. There is no TIS value reported for H2RG-122. Values for H1RG-161 and H2RG-122 are lower than for the other two detectors. TIS used different processing techniques on these detectors, specifically designed to lower dark current.

H1RG-125 has an aluminum optical blocking filter on one half of the detector and no filter on the other half. During a previous round of testing, in 2008, the detector was accidentally exposed to bright optical light while powered up [10]. The dark current measurements for this detector are differentiated between the two halves. Table 4 shows that at 150 K the dark current for the filtered and unfiltered halves are consistent within the measurements uncertainty, 0.280 ± 0.080 (e$^-$ sec$^{-1}$ pixel$^{-1}$, filtered) and 0.230 ± 0.023 (e$^-$ sec$^{-1}$ pixel$^{-1}$, unfiltered). The filtered results are also consistent with the TIS measurements, 0.214 (e$^-$ sec$^{-1}$ pixel$^{-1}$), made before the detector was used at PSU. The detector dark current is uniform across the entire chip at 150 K. Figure 12 (upper left panel) shows the H1RG-125 dark current as a function of temperature. The dark current increases much faster with temperature on the unfiltered side.

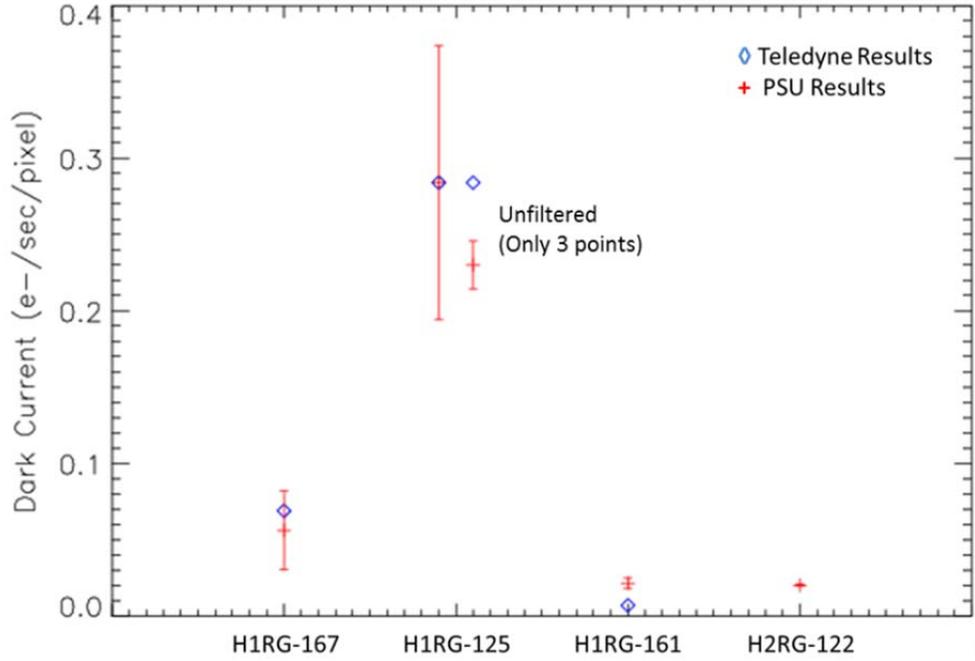

Figure 10. Dark current measurements at 150 K made by TIS and PSU. Two results are given for H1RG-125, filtered and unfiltered sides of the detector (filtered on the right, unfiltered on the left). NOTE: There is not a TIS measurement available for H2RG-122 and error bars are not known for the TIS measurements.

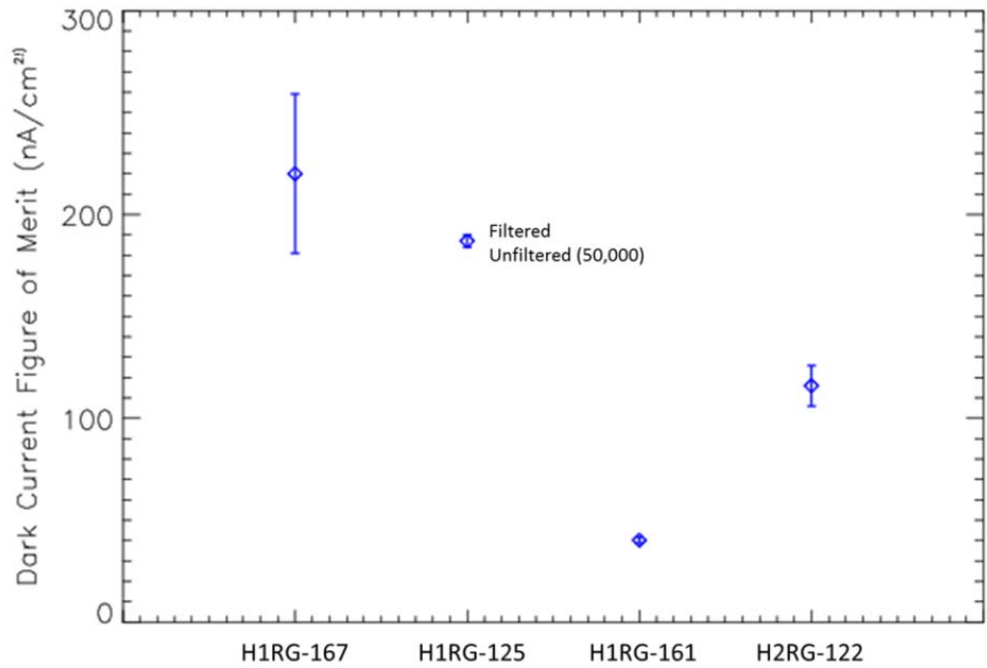

Figure 11. Dark current figure of merit at 293 K from fits described in Sec. 3.3 and reported in Table 4 as (e$^-$ sec$^{-1}$ pixel$^{-1}$). The data point for H1RG-125 shown in the plot is for the filtered side; the unfiltered side had a measured $DC_{FM}$ of approximately 50,000 (nA cm$^{-2}$) at 293 K.

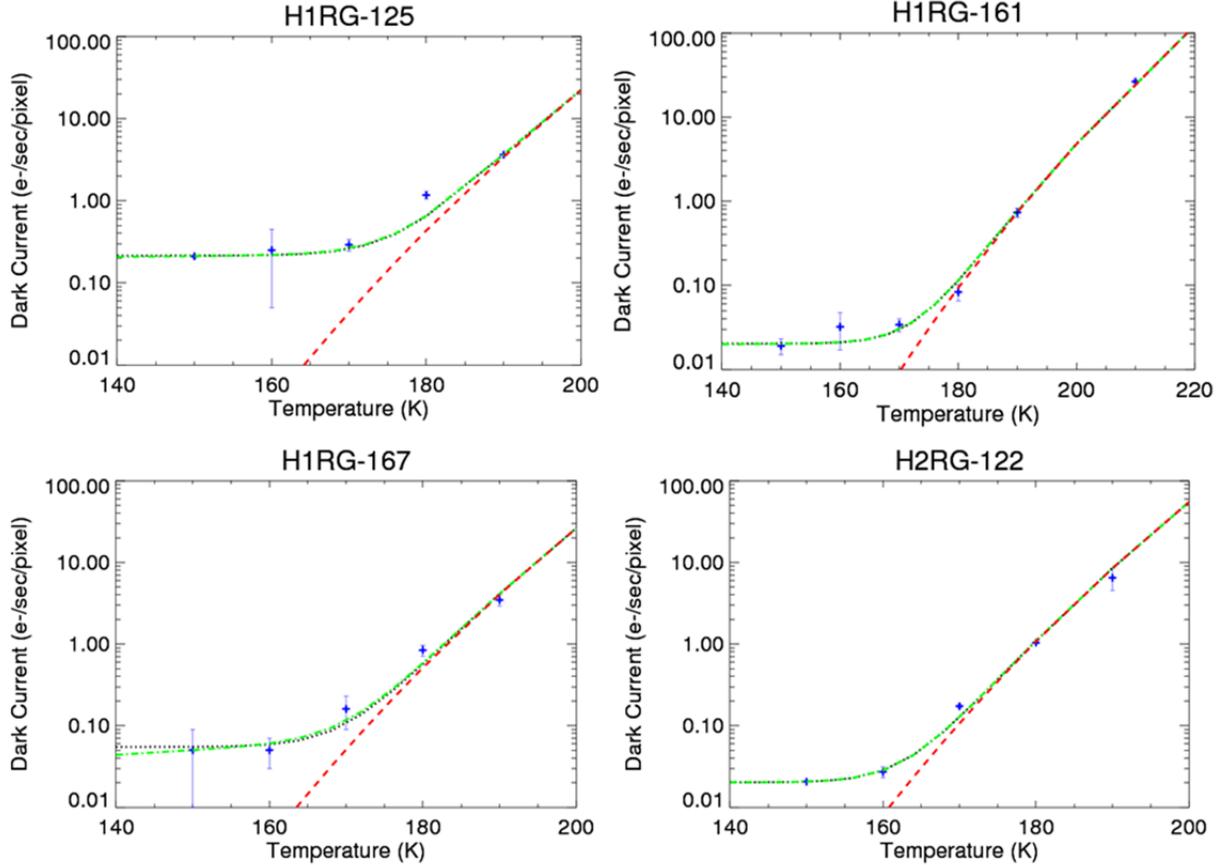

Figure 12. Dark current measurements at different temperatures for our four detectors. The data are fit with a theoretical dark current function, Equation 1 (Red Dash), plus a constant (Black Dot) and an additional power law (Green Dot-Dash) described in more detail in Sec. 3.3.

Plotting dark current versus temperature shows that each of the four detectors fits the function given by Equation 1 above approximately 180 K, Figure 12. At temperatures below this however, there is another contribution that needs to be included. To fit this portion of the data we attempted multiple models, including Equation 1 with multiple $D_{FM}$, a linear function and a constant. The constant gave the best results; given the small number of data points it is difficult to draw detailed conclusions from this fit but there appear to be two different dark current distributions. Using our fit for each detector we report the dark current figure of merit ($D_{FM}$)–the dark current calculated at 293 K–in Table 4 (e$^-$ sec$^{-1}$ pixel$^{-1}$) and Figure 11 (nA cm$^{-2}$).

## 4. CONCLUSIONS

We have reported on the characterization of TIS H1RG detectors along with a specially designed H2RG detector. Measurements of energy resolution at multiple energies were made. At 150 K, the resolution of a 5.9 keV line was found to be as good as 248 eV (4.2 %) for the best detector [10] and 543 eV (9.2 %) for the worst, with the effects from IPC contributing to the high values. H2RG-122 has IPC characteristics good enough to treat the detector like a CCD when measuring energy resolution but the improvement in energy resolution is limited by the fact that the detector is an engineering model with high read noise and response variations across the detector. Using only single and split pixel events we measure an energy resolution of 443 eV at 5.9 keV (7.5 %), but this improves to 369 eV (6.3 %) when only using single pixel events. The resolution at other energies and temperatures are reported in Table 3.

Read noise was measured and reported in Table 4. We measured a low value of 7.1 e- (RMS). While not at the sub-e- level of the best CCDs, these values show continuing improvement in HCD noise with promise of further improvement.

Interpixel capacitance crosstalk (IPC) was measured for each detector using an average of single pixel events from an $^{55}$Fe source. We report the full results for each detector in Sec. 3.2. The H1RG detectors have IPC upper limits of 4.0 – 5.5 % (up/down) and 8.7 – 9.7 % (left/right), showing a clear asymmetry in the IPC between pixels of the same row and pixels in the same column. The most likely explanation for IPC is that it is caused by capacitive coupling between neighboring absorber and/or ROIC pixels. The cause of the asymmetry is, however, unclear. The H2RG detector was expected to have lower IPC due to the increased distance between ROIC pixels, thus reducing the capacitive coupling effect. The H2RG-122 detector had an IPC of 1.8 ± 1.0 % and no measurable asymmetry, a clear improvement over the IPC for 18 $\mu$m pitch H1RGs. The next goal will be to reduce the size of these pixels while maintaining a low IPC.

We have measured dark current at a range of temperatures from 150 – 210 K. The results at 150 K varied from a high value of 0.280±0.080 (e$^-$ sec$^{-1}$ pixel$^{-1}$) for H1RG-125 to a low value of 0.020 ± 0.001(e$^-$ sec$^{-1}$ pixel$^{-1}$) for H2RG-122. The dark current as a function of temperature data can be well-fit by Equation 1 plus a constant, although with the low number of data points, other models are also viable, such as Equation 1 plus a power law component. The need for this additional function indicates the presence of two sources of dark current, where one dominates at temperatures above ~180 K and the other below. The dark current below 170 K is insignificant compared to the read noise, thus the detectors could potentially be run at temperatures as high as 170 K without degrading the energy resolution.

The energy resolution, read noise, interpixel capacitance and dark current measurements made for this project have shown the continuing promise of Hybrid CMOS detectors (HCDs) for the future of X-ray astronomy. The strong positives of HCD technology, high radiation hardness, fast readout, and windowing capability, should allow HCDs to pave the way for future X-ray telescopes to accomplish science goals like observing the gravitational effects around distant black holes and understanding the nature of ultra-dense material found in neutron stars.

For missions that require radiation hardness, low power, and fast readout, without the requirement for very low read noise or Fano-limited energy resolution, these hybrid silicon CMOS detectors are already an excellent choice (e.g. JANUS [15]). For missions requiring these advantages, along with less than 4 e$^-$ read noise and small pixels (e.g. SMART-X [1]), these detectors offer promise which can be realized with continued development.

## ACKNOWLEDGEMENT

We gratefully acknowledge Teledyne Imaging Systems, particularly Mark Farris, James Beletic, and Yibin Bai, for providing useful comments and for loaning us the H2RG-122 detector. This work was supported by NASA grants NNG05WC10G, NNX08AI64G, and NNX11AF98G.